\documentclass{article}
\usepackage{makeidx,epsfig}
\usepackage[round]{natbib}
\usepackage{setspace,graphicx}
\usepackage{Generic}
\makeindex

\newcommand{\be}{\begin{equation}}
\newcommand{\ee}{\end{equation}}

\author{Brian Skinner \\ Massachusetts Institute of Technology, Cambridge, MA 02139  USA
\and  Matthew Goldman \\ Microsoft Corporation, Redmond, WA 98052  USA}

\date{\today}

\title{Optimal Strategy in Basketball}

\begin{document}

\maketitle

\begin{abstract}

This book chapter reviews some of the major principles associated with optimal strategy in basketball.  In particular, we consider the principles of allocative efficiency (optimal allocation of shots between offensive options), dynamic efficiency (optimal shot selection in the face of pressure from the shot clock), and the risk/reward tradeoff (strategic manipulation of outcome variance).  For each principle, we provide a simple example of a strategic problem and show how it can be described analytically.  We then review general analytical results and provide an overview of existing statistical studies.  A number of open challenges in basketball analysis are highlighted.

\end{abstract}

\section{Introduction}

The game of basketball, like just about any other game, can be described as a sequence of random events, with the probabilities of different outcomes determined by the skill levels of the players involved.  More specifically, during each offensive possession of a basketball game, the team with the ball attempts to score, and this attempt results in some number of points ranging between 0 and 3 (or, rarely, 4).  At the end of the game, what matters is not the absolute score of a given team but the differential score, defined as $\Delta = (\textrm{total points scored by team A}) - (\textrm{total points scored by team B})$.  The goal of team A is for the game to end with $\Delta > 0$.  As such, this differential score is the primary random variable for describing a basketball game in the statistical sense.  Understanding its expected value and variance, and how basketball teams can influence them through strategic decisions, is the main focus of this chapter.

Despite some important exceptions (which will be discussed in detail in the Conclusion section, and which do not fundamentally change the strategic considerations of this chapter), empirical studies of scoring in basketball suggest that scoring events can be modeled with a fairly high degree of accuracy as independent and identically distributed (IID). This statistical independence implies that, as the game progresses, the differential score undergoes a biased random walk.  Its random fluctuations are therefore nearly identical to those of Brownian motion \citep{gabel_random_2012, clauset_safe_2015}, while the net drift of the score is determined by the skill levels of the two teams and by the scoring strategies they employ.


In the limit where there are many possessions remaining in the game, one can invoke the central limit theorem, which says that regardless of the process used by each team to score during a given possession, the probability density $P(\Delta)$ for many possessions is described by a Gaussian distribution:
\be
P(\Delta) = \frac{1}{\sqrt{2 \pi \sigma^2}} e^{-(\Delta-\mu)^2/2\sigma^2}. \label{eq:gaussian}
\ee 
Here, $\mu$ is the expected value of the differential score and $\sigma^2$ is its variance. 

The intent of strategic decisions in basketball, at the level of a single game, is to maximize the probability of victory, $\mathcal{P} = \int_0^{\infty} P(\Delta) d\Delta$, i.e., the probability of ending the game with $\Delta > 0$.  
In most cases, maximization of $\mathcal{P}$ is as straightforward as trying to maximize the expected number of points scored by the team, and to minimize the expected number of points scored by the opponent.  As we discuss below, however, this is not always the case, particularly in late-game situations.

The remainder of this chapter is dedicated to discussing the principles associated with optimal strategy in basketball -- that is, with the maximization of $\mathcal{P}$.  Of course, it should go without saying that trying to maximize the variable $\mathcal{P}$ is not as straightforward as your typical math problem.  Generally speaking, it is as tricky and nuanced as basketball itself, and requires one to think about the principles of teamwork, leadership, and performance under pressure.  Nonetheless, our goal in this chapter is to review how these all-important concepts can be described quantitatively, and how they ultimately relate back to a mathematical maximization of $\mathcal{P}$.

We focus our discussion in this chapter around three general principles associated with optimal strategy in basketball.  We refer to these principles as allocative efficiency (Sec.\ \ref{sec:allocative}), dynamic efficiency (Sec.\ \ref{sec:dynamic}), and the risk/reward tradeoff (Sec.\ \ref{sec:risk}).  For each of these three principles, we provide an illustrative example and present an overview of general analytical results.  We then provide some preliminary discussion of how these principles can be combined with data from real basketball games to evaluate and improve the decision making of basketball teams.

\section{Allocative Efficiency}
\label{sec:allocative}

When planning its offensive strategy, the primary decision faced by a basketball team is this: which shots should the team take, and which players should take them?  In more general language, this question becomes: how does the team optimally allocate its shot attempts between different offensive options? Or, from the perspective of players in the flow of the game: which shooting opportunities are \textit{good enough} to be taken, and which should be passed up?

The most na\"{i}ve approach to this optimization problem would be to simply determine which play (or player) provides the highest expected return in terms of points scored, and then to run that play exclusively. For example, in the 2014-15 season of the National Basketball Association (NBA), Kyle Korver led the league with a true shooting percentage of $0.699$.\footnote{Data from http://www.basketball-reference.com/} This statistic essentially means \citep{kubatko_starting_2007} that for every shot Kyle Korver took, his team, the Atlanta Hawks, scored $2\times 0.699 = 1.398$ points,\footnote{This number does not include points the Hawks might score after rebounding Kyle Korver's missed shots. It is likely that the Hawks averaged more than 1.5 points in total on possessions where he took a shot.} as compared to only $1.089$ points on their average possession. Despite his high efficiency, however, Korver only attempted 8.0 shots per game in the 2014-15 NBA season (less than $10\%$ of his team's total). Clearly, the Hawks are failing to adhere to such a simple model of optimality. But should Kyle Korver really be taking many more shots?

To answer this question, consider what it would mean for Korver to take a significantly larger proportion of his team's shots.  A substantial increase in Korver's shooting rate could be accomplished in two different ways.  First, the Hawks could run many more plays that are designed to end with a Korver shot.  In doing so, however, the Hawks would risk becoming highly predictable offensively -- that is, the defense would learn to focus their attention on Korver, and his effectiveness as an offensive option would plummet.  A second option for the Hawks would be to run their usual offense, but to have Korver simply increase his willingness to shoot, taking contested shots that he might usually have passed on. By definition, these \emph{marginal} shots are not ones that Korver would have taken previously, and they will (presumably) result in fewer points on average.

In either scenario, it is reasonable to expect that the average return of a Korver shot decreases as Korver's offensive load is increased.  In more general terms, this example highlights the crucial idea that the effectiveness of a particular play (or player) is not adequately described by a single number.  Instead, a play should be characterized by a \emph{function} that describes how the expected return of the play declines with increased use. 

In the seminal work by \citet{oliver_basketball_2004}, this relationship between efficiency and usage is called a ``skill curve", which we denote here by $f(p)$.  Oliver proposed that this relation could be defined such that $f$ represents the average number of points scored by a given player (per possession used) as a function of the proportion $p$ of possessions used by that player while he/she is on the court. Under this definition, the relation $f(p)$ is descriptive of the player's skill level: highly skilled players have a skill curve $f(p)$ that declines very little with increased offensive load $p$, while less skilled players have a curve $f(p)$ with a more strongly negative slope.\footnote{Indeed, this pattern was documented in \citet{goldman_allocative_2011}, which found that guards tend to have flatter usage curves than forwards and centers, and that higher scoring players have flatter usage curves more generally.}  The first attempts to measure skill curves involved tallying the number of points scored by a given player per possession, $f$, and the fraction of possessions used by that player, $p$, for all games within a particular time period.  Estimates for $f(p)$ were then produced by sorting games into bins by their value of $p$ and making a simple average of $f$ for each bin.  

In this chapter, on the other hand, we adopt a more generalized definition of the relation $f(p)$.  In particular, we take $p$ to be a fraction that represents the number of times that a particular play is used (i.e., the number of times that that play is used normalized by the total number of possessions), and we take $f$ to be the average number of points scored per possession when that play is run.  We generally refer to $f(p)$ as a ``usage curve", which can refer to a specific play rather than to the aggregate effectiveness of all shots taken by a particular player.


While a quantitative determination of the function $f(p)$ for a given play or player can be a difficult statistical problem, the idea that the effectiveness of a play declines with increased usage is a fairly intuitive one.  Among players and coaches, this idea is often expressed by saying that the team needs to ``keep the defense honest,'' or to ``take what the defense gives them."  But using usage curve relations to reach quantitative decisions about optimal strategy requires more careful consideration.

To get a sense of how this optimization process works, imagine a very simplified hypothetical example of a strategic decision in basketball, in which a team is attempting to optimize the frequency $p$ with which it takes corner three-point shots (i.e., $p$ is the fraction of all shots that are corner $3$'s).  Suppose it is known that the average number of points scored per attempted corner three, $f$, follows the relation $f(p) = 2(1-p)$.  In other words, in this example the corner three produces as much as $2$ points on average when attempted very rarely (i.e., the shot is made $67\%$ of the time), and produces close to zero points if it is attempted every time down the court.  All other offensive options, taken together, are assumed to produce a constant $1.0$ points per possession.  What is the optimal use of the corner three by this offense?

\begin{figure}[htb]
\centering
\includegraphics[width=0.7 \textwidth]{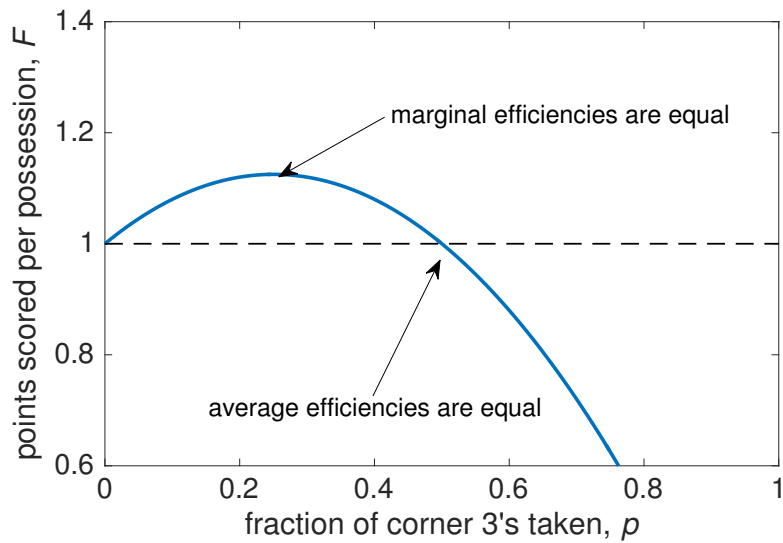}
\caption{Optimal usage of the corner $3$-pt shot in the simplified example of this section.  The solid line shows the average number of points scored per shot when a fraction $p$ of the team's shots are corner $3$s.  The dashed line shows the effectiveness of all non-corner-$3$ options.  Choosing $p = 0.5$ equates the average effectiveness of all offensive options, but $p = 0.25$ equates their marginal effectiveness and provides the optimal solution.}
\label{fig:corner3}
\end{figure}

The answer to this question can be found in a simple way by writing down a function $F$ that describes the expected total number of points scored as a function of $p$:
\be 
F = p \times f(p) + (1-p)\times 1.0.
\nonumber
\ee 
In this equation, the first term corresponds to the number of points scored from corner $3$s, and the second term is the number of points scored by all other shots.  The optimal value of $p$ is the one that maximizes $F$, and it can be found by taking the derivative of $F$ with respect to $p$ and equating it to zero:
\be 
\frac{dF}{dp} = \frac{d}{dp}\left(p f(p)\right) - 1.0 = 0.
\ee 
Solving this equation gives $p = p_\textrm{opt} = 0.25$.  That is, in this example the optimal proportion of corner $3$'s is 25\%.  This is illustrated in Fig.\ \ref{fig:corner3}.

It is important to note that in this optimal strategy, the corner three pointer still produces significantly more points on average than other, non-three point options ($1.5$ points as compared to $1.0$ points).  Noticing this disparity, one would be tempted to suggest that more corner $3$'s should be taken, since they produce a larger average return.  Such a conclusion, however, is a mistake, and it arises from focusing on the \emph{average} shot quality $f(p)$ rather than the \emph{marginal} shot quality $d(p f(p))/dp$.  In the correct solution, it is the marginal values of the two options that become equal, and not the average values.

In fact, one can generalize this result into a principle that we refer to as the \emph{allocative efficiency criterion}:
\begin{center}
{\bf \textit{In an optimal strategy, all offensive options have the same marginal efficiency.}}
\end{center}

Formally, the allocative efficiency criterion can be written as follows.  Suppose that different plays $1, 2, 3, ...$ are characterized by functions $f_1(p_1)$, $f_2(p_2)$, $f_3(p_3)$ ..., which describe the average number of points scored per possession by each play as a function of the frequencies $p_1$, $p_2$, $p_3$ ... with which the plays are run.  The play frequencies are defined such that $p_1 + p_2 + p_3 + ... = 1$.  The choice of different play frequencies defines an offensive strategy.  The expected number of points scored per possession for a given strategy is given by
\be 
F = p_1 f_1(p_1) + p_2 f_2(p_2) + p_3 f_3(p_3) + ... .
\label{eq:Falloc}
\ee 
This function is maximized when
\be 
\frac{d(p_1 f_1(p_1))}{d p_1} = \frac{d(p_2 f_2(p_2))}{d p_2} = \frac{d(p_3 f_3(p_3))}{d p_3} = ...
\label{eq:allocative}
\ee
Since the quantity $d(p_i f_i(p_i))/dp_i$ represents the marginal efficiency of play $i$, equation (\ref{eq:allocative}) is the quantitative form of the allocative efficiency criterion.

If the usage curves $f_i(p_i)$ are all known, then in principle the optimal strategy can be found by solving the set of equations in  (\ref{eq:allocative}).  In the case where all functions $f(p)$ are linear, so that $f_i(p_i) = \alpha_i - \beta_i p_i$, there is a relatively simple analytical solution.  (Here $\alpha_i$ and $\beta_i$ are constants and the index $i$ runs from $1$ to $N$, where $N$ is the total number of plays being considered by the offense.)  In this case the optimal frequency of play $i$ is given by \citep{skinner_price_2010}
\be 
p_{i,\textrm{opt}} = \frac{\alpha_i + \lambda}{2 \beta_i},
\label{eq:xopt}
\ee 
where $\lambda$ is a constant given by\footnote{It is possible that, under certain somewhat unrealistic conditions, equation (\ref{eq:xopt}) can give negative solutions for the frequency $p_j$ of some play $j$.  In this case one should set $p_j = 0$ and re-solve for the remaining $p$'s.}
\be 
\lambda = \frac{2 - \sum_{i = 1}^N \alpha_i/\beta_i}{\sum_{i = 1}^N 1/\beta_i}.
\ee
In cases where the functions $f_i(p_i)$ are not linear, one can find the optimum strategy by using a numerical search for the maximum of equation (\ref{eq:Falloc}).

The topic of shot allocation in basketball has been the subject of a significant number of studies during the past two decades, with most studies focusing on the choice between two-point and three-point shots.  For example, in \citet{vollmer_application_2000, alferink_generality_2009, neiman_spatial_2014}, the proportion of three-point shots was compared to predictions from the so-called ``matching law" from psychology.  The matching law, generally speaking, holds that there is a direct relation between the rate of incidence of a behavior and its rate of reinforcement.  In the context of $2$-point and $3$-point shots, the matching law predicts that the proportion of attempted $3$'s should match the proportion of points scored by such shots, and this law seems to closely match the shooting habits of NBA players.  Of course, adherence to this matching law does not necessarily indicate optimal shot selection.  Other studies have examined the $2$-point/$3$-point choice through the lens of reinforcement learning, and have shown that NBA players tend to over-generalize from previous successes \citep{neiman_reinforcement_2011} and to use only the $2$-point/$3$-point distinction as the primary characteristic for generalization in learning \citep{neiman_spatial_2014}. Thus, it may be the psychology of reinforcement, rather than a detailed understanding of optimal behavior, that drives shot selection in basketball players.

Some previous studies have examined the question of optimal play selection in soccer, football, and baseball through the paradigm of the ``minimax" rule in game theory \citep{palacios-huerta_professionals_2003,kovash_professionals_2009}.  The minimax rule generally dictates that an optimal game-theoretical strategy (say, for the defense) is the one that minimizes the worst case scenario (the number of points given up).  These studies, however, evaluated adherence to the minimax strategy only by examining the \emph{average} return from different offensive options, rather than the marginal return.  Such studies are therefore unlikely to be able to gauge whether basketball teams are exhibiting optimal allocation between plays.

Finally, a number of studies have sought to quantify or refine the definition of ``shot value" by either examining the spatial structure behind patterns of shot effectiveness \citep{reich_spatial_2006, cervone_multiresolution_2014, miller_factorized_2014, chang_quantifying_2014, shortridge_creating_2014} or by attempting to describe the offense as a branching network of options \citep{fewell_basketball_2012, skinner_method_2015}.  
In \citet{goldman_optimal_2014, goldman_allocative_2011}, a method is presented for approximating the usage curve of a given player (or a given unit) by examining the shooting efficacy and the shooting rate as a function of the shot clock time.  In particular, these authors show that one can estimate how an offensive option performs under the pressure of increased usage by observing its performance under the pressure of a dwindling shot clock.  However, constructing a more robust method of measuring usage curves for specific plays/players remains a prominent challenge.

\section{Dynamic Efficiency}
\label{sec:dynamic}

In addition to the dynamics associated with allocation that are outlined in the previous section, basketball strategy also has an important time element.  Specifically, in most competitive basketball games, the teams' decision making is constrained by a shot clock.  As the clock winds down, the expected return of a given possession declines, and as a team becomes increasingly desperate it must reconsider its previous unwillingness to take certain lower quality shots.  In this way, optimal decision making in basketball is by nature \emph{dynamic}, and depends on the time remaining on the shot clock.

In this sense, the problem of shot selection in basketball falls into the class of ``optimal stopping problems", wherein one is concerned with trying to choose the optimal moment to perform a specific action in the face of finite time and uncertainty about future events.\footnote{The most famous optimal stopping problem is the so-called ``secretary problem", which imagines the process of choosing a secretary from a long line of applicants under the constraint that a candidate must be hired or dismissed immediately following the interview.}  Similarly, in basketball the player with the ball must decide at the instant of a shot opportunity whether the available shot is good enough to take, or whether it is smarter to pass up the shot and wait for a better opportunity.  This decision necessarily involves weighing the expected return of the remainder of the possession.  In mathematics and economics, such decisions are commonly described analytically by a Bellmann equation.  As we show below, writing down an analog of the Bellmann equation for the shot selection process brings us to a general criterion for dynamic efficiency.

As a way of introducing the central ideas associated with dynamic efficiency, consider the following hypothetical game.  Imagine that you are challenged to make a basketball shot, and that the difficulty of the shot is chosen by spinning a roulette-style wheel.  (For example, the wheel could dictate the distance from which the shot has to be taken.)  After each spin of the wheel, you are given two choices: you can either attempt the shot, or you can choose to spin the wheel again.  You are given a finite number $N$ of wheel spins (which we'll call ``shot opportunities"), and the game ends when you have either taken a shot or run out of shot opportunities.  The crucial strategic question then becomes: how do you decide when a shot is good enough to take, and when you should spin again?  What will be the winning percentage associated with the optimal strategy?

When $N = 1$, meaning that you are only given one shot opportunity, the answers to both of these questions are simple.  Since you only get one shot, you should take it.  Your winning percentage will then be equal to the expectation value of your shooting percentage over all possible distances written on the wheel.  In other words, if one defines the ``shot quality" $\eta$ for each possible distance as the probability that you will make the shot from that distance, then the chance that you will win the game is $V_1 = E[\eta]$ when $N = 1$.  

When $N = 2$, on the other hand, you can afford to be more selective about your first shot.  In particular, if the first available shot has a quality $\eta_2$ that is smaller than $V_1$, then it makes sense to pass up that first shot and spin the wheel again.  This strategy will produce an expected winning probability $V_2$ that is strictly larger than $V_1$.  Similarly, when $N = 3$, the first shot should be taken only when its quality exceeds the combined value $V_2$ of the next two shots, taken together.

This line of inductive reasoning leads to a recurrence equation for the winning probability $V_N$, based on the principle that when $N$ shots are remaining, the first shot opportunity should be taken only when its quality $\eta$ exceeds $V_{N-1}$. This recurrence relation was studied analytically by \citet{skinner_problem_2012}. The numeric values of the sequence $V_N$ depend on the probability distribution for $\eta$ at each opportunity (i.e., which distances are written on the wheel, and how good you are at shooting from those distances).

Of course, this hypothetical game is significantly different from the real shot selection problem faced by basketball teams.  For example, in basketball the shot clock runs continuously rather than in discrete turns, and there is no reason to think that the distribution of shot quality $P(\eta)$ will be independent of the shot clock time.  Nonetheless, the line of inductive reasoning used above can be extended to derive a general criterion for dynamic efficiency in basketball.

Specifically, let us first divide up the full length of the shot clock (usually, either 24 or 35 seconds) into $1$-second intervals.  Let $\eta_1$ be a random variable that describes the quality of a shot opportunity available to a team when there is only one second remaining on the shot clock.  ($\eta_1$ can be defined as the probability that the shot will be made multiplied by the point value of the shot.)  When only $1$ second is remaining on the clock, the team with the ball must shoot regardless of the value of $\eta_1$, and as a consequence the expected number of points scored by the shot is just $V_1 = E[\eta_1]$.   

When there are $t = 2$ seconds remaining, on the other hand, the team can afford to be slightly more selective.  In particular, a shot should be taken immediately only if its quality $\eta_2$ exceeds $V_1$.  The average return from this strategy, in terms of the number of points scored, is $V_2 = P(\eta_2 \leq V_1)\cdot V_1 + P(\eta_2>V_1) \cdot E[\eta_2 | \eta_2>V_1]$, which is the weighted average of the value the team gets when it waits until $t=1$ (the first term) and the average quality of the shot they take in instances when the team shoots at $t=2$ (the second term).

This line of thinking can be generalized for arbitrary time $t$ into the following set of Bellmann-like equations:
\begin{eqnarray}
V_t & = & P(\eta_t \leq c_t) \cdot V_{t-1} + P(\eta_t > c_t) \cdot E[\eta_t | \eta_t > c_t]  \nonumber \\
c_{t} & = & V_{t-1} .
\label{eq:reservation}
\end{eqnarray}
The quantity $V_{t}$ can be called the \emph{value} of the possession when $t$ seconds are remaining, and $c_t$ is the \emph{reservation threshold} for shooting with $t$ seconds left.  The optimal strategy is defined by shooting if and only if the quality $\eta_t$ of a given shot opportunity exceeds $c_t$, with $c_t$ given by Eqs.\ (\ref{eq:reservation}).

In conceptual terms, one can state the conclusion of Eqs.\ (\ref{eq:reservation}) as the following \emph{dynamic efficiency criterion}:
\begin{center}
{\bf \textit{Shots should be taken only when their expected return exceeds the average value of continuing the possession.}}
\end{center}

Empirically testing whether teams adhere to dynamic efficiency is a complicated statistical problem.  While the value $V_t$ of continuing a possession can be readily measured from data, the reservation threshold $c_t$ is not directly observed.   Instead, we can only see average efficiencies from \textit{all} shots taken with $t$ seconds remaining. That is, we see $e_t = E(\eta_t | \eta_t>c_t)$, which is a number strictly greater than $c_t$. Another way to say this is that we are interested in the very worst shot a team is willing to take with $t$ seconds remaining on the shot clock, but can only observe the outcomes of the average of all shots taken with $t$ seconds remaining. This average will include some high-value, wide open shots that were easy decisions for the offensive team, and as such is not representative of the team's reservation threshold.  One can, of course, apply a simple check for dynamic inefficiency by measuring whether $e_t > V_{t-1}$ for all values of the shot clock time $t$.  If this condition is violated, then the team is overly willing to settle for low quality shots at time $t$, since $c_t = V_{t-1}$ represents the lowest quality shot that a team should take and therefore it should never exceed $e_t$.  However, $e_t > V_{t-1}$ does not by itself guarantee optimal dynamic efficiency.  

\citet{goldman_optimal_2014} studied the question of dynamic efficiency by estimating a structural model in which the time remaining on the shot clock is used as an instrumental variable for shot selection. In this way, they were able to model the marginal (or worst) shot a team is willing to take and compare its quality to the average number of points returned from continuing the possession. Figure \ref{fig:pointvalues} below shows how both the average value of a team's possession and the reservation threshold decline in perfect unison as the shot clock ticks toward zero (from left to right), thus demonstrating impressive adherence to dynamic efficiency.

\begin{figure}[htb]
	\centering
	\includegraphics[width=0.8 \textwidth]{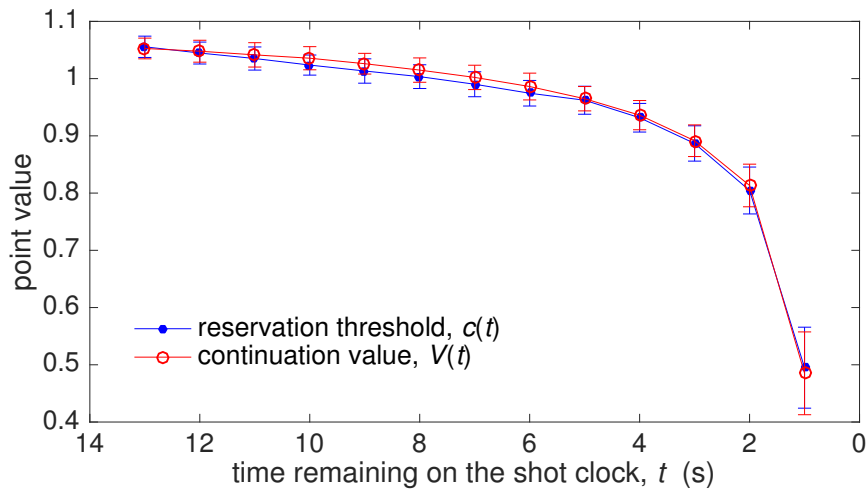}
	\caption{Average continuation values (red open circles) and reservation thresholds (blue points) as a function of the shot clock time for the most common lineups in the NBA between 2006 and 2012. Data taken from \citet{goldman_optimal_2014}.}
	\label{fig:pointvalues}
\end{figure}

\section{Risk and Reward}
\label{sec:risk}

In the preceding two sections, we have essentially approached the question of optimal strategy as equivalent to the question of ``how does a team maximize the expected differential score?"  In other words, we have only discussed the process of trying to maximize the expected number of points scored by the offense per possession.  But, in fact, this is not the same question as ``how does a team maximize its chance of winning?"  A true optimal strategy considers not just the expected outcome, but the full distribution of possible outcomes.

To see the difference, consider the following simple example.  Imagine that a friend challenges you to a shooting contest.  The rules of the game are that you each get to take $10$ shots, either from the free throw line or from the three point line.  The free throw shots are worth $2$ points each, and the three-point shots are worth $3$.  Suppose, further, that you know yourself to be a $60\%$ free throw shooter and a $40\%$ three-point shooter.  Which shots should you choose to take?

As it turns out, the answer to that question depends entirely on whether your friend is a better shooter than you are.

In this example, the shooting percentages are such that both strategies (taking either all $2$'s or all $3$'s) have the same expected outcome: $12$ points scored.  But the two strategies differ in the variance of the outcome.  In particular, the three point-shooting strategy offers both a higher ``best case scenario", and a lower ``worst case scenario".  So the decision of whether to pursue that strategy depends on whether you expect that beating your friend is a likely or an unlikely event.  If winning is likely, then you should go with the more conservative $2$-point shooting strategy.  If it is unlikely, then you should choose the more risky $3$-point shooting strategy.

More formally, the ``riskiness" of a strategy can be quantified by the variance $\sigma^2$ in the total number of points scored.  Under the assumption that individual possessions are IID, the variance is simply the sum of the variances $\sigma_i^2$ of each possession.  This single-possession variance is given by
\be 
\sigma_i^2 = \sum_{j} P_j \cdot (f(j) - \mu_i)^2,
\ee
where $j$ indexes the set of possible outcomes of the possession, $P_j$ is the probability of each outcome, $f(j)$ is the number of points scored in outcome $j$, and
\be 
\mu_i = \sum_{j} P_j \cdot f(j)  
\ee
is the expectation value of the number of points scored in possession $i$.  So, for example, in the simple example above, the expected number of points scored for a single $2$-point shot, $\mu_2 = 1.2$, is the same as the expected number of points scored for a $3$-point shot.  However, the variance for a single $2$-point shot, $\sigma^2_2 = (\textrm{probability of making the shot}) \times (2 - \mu_2)^2 + (\textrm{probability of missing the shot}) \times (0 - \mu_2)^2 = 0.6 \times (2-1.2)^2 + 0.4 \times (0-1.2)^2 = 0.96$\,pts$^2$, is substantially smaller than the variance for a single $3$-point shot, $\sigma^2_3 = 0.4 \times (3 - 1.2)^2 + 0.6 \times (0 - 1.2)^2 = 2.16$\,pts$^2$.  

In the example above, a strategy of taking ten $2$-pointers has a variance $\sigma^2 = 10 \times 0.96 = 9.6$\,pts$^2$ while a strategy of taking $10$ $3$-pointers has a variance $\sigma^2 = 21.6$\,pts$^2$.  Thus, one can say that the $2$-point strategy has a range of outcomes $\mu \pm \sigma = 12 \pm 3.1$ pts, while the $3$-point strategy has a range $12 \pm 4.65$ pts.  So, for example, if you expect that your friend is going to score $15$ points, then it makes more sense to shoot $3$'s.  If you only expect your friend to score $8$ points, on the other hand, it's smarter to go with the more conservative $2$-point shots.

This example is relatively straightforward, since it involves choosing between two strategies with the same expected return.  Choosing the correct strategy is therefore as simple as deciding whether large variance is a benefit (since it increases the probability of an unlikely upset, if you are the underdog) or a drawback (for the same reason, if you are the favorite).  In most realistic scenarios, however, one must choose between strategies that do not have the same expected return.  In such cases the ideal strategy may involve sacrificing from the expected number of points scored in order to increase the variance (if you are the underdog) or reduce it (if you are the favorite).

\begin{figure}[htb]
\centering
\includegraphics[width=0.7 \textwidth]{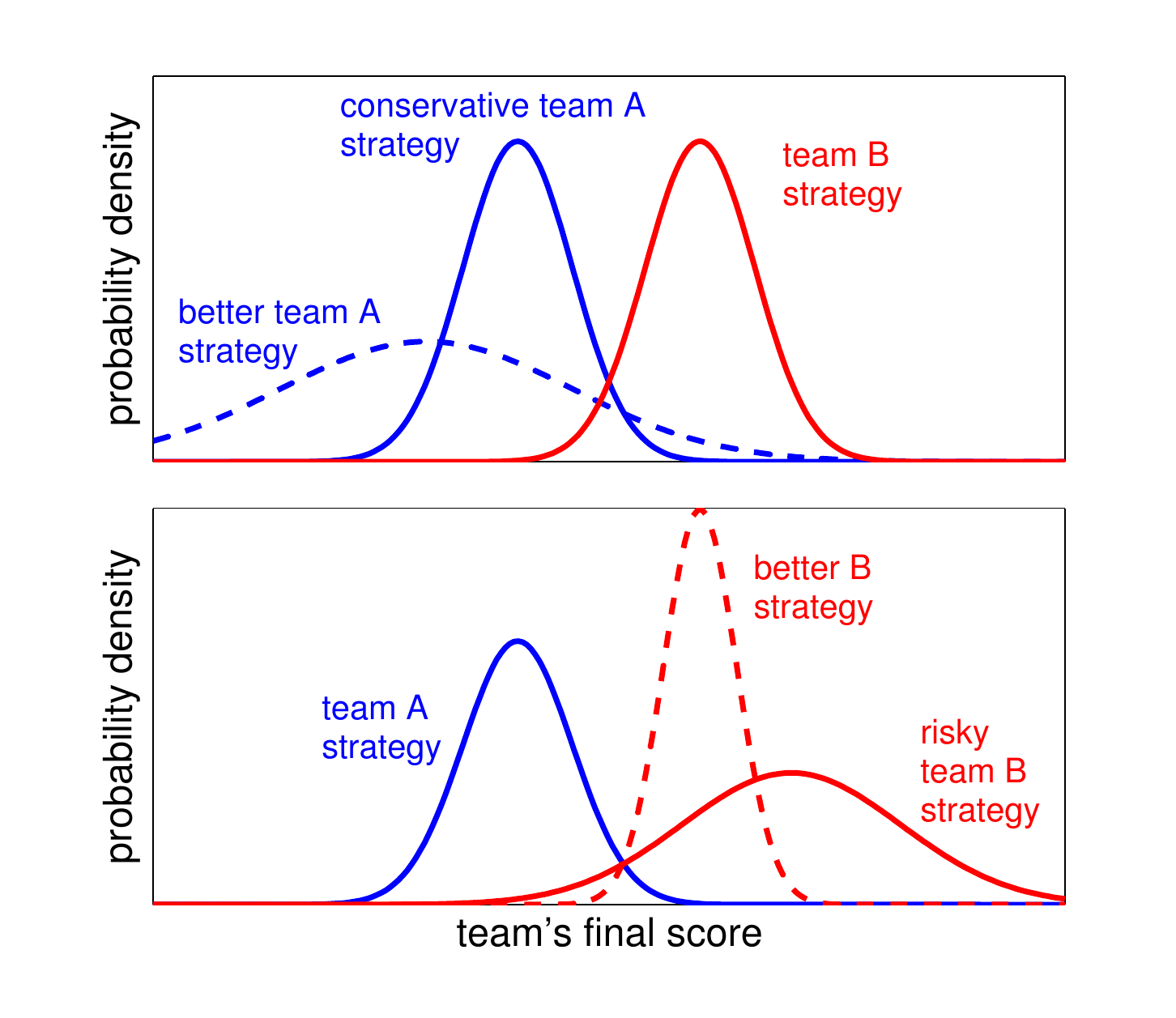}
\caption{Schematic picture of the distribution of final scores for two competing teams. Above: The underdog team (blue) improves its chance of winning by pursuing the risky strategy corresponding to the dashed line rather than the conservative strategy corresponding to the solid line, even though this lowers the expected final score. Below: Conversely, a favored team can improve its chance of winning by using a conservative strategy (dashed line) rather than a risky strategy (solid line), even if this lowers its expected final score.}
\label{fig:distributions}
\end{figure}

Graphically, one can say that the strategy adopted by each team is characterized by a distribution of possible final scores.  For an underdog team, it is in the team's best interest to choose a strategy that maximizes the overlap between the two teams' distributions.  The favored team, on the other hand, should choose the strategy that minimizes the overlap between the two distributions.  This is shown schematically in Fig.\ \ref{fig:distributions}.

At the most general level, one can describe the problem of optimal strategy in the following way.  Let the final number of points $s_A$ scored by team $A$ be described by a probability distribution $P_A(s_A)$.  Let $P_B(s_B)$ similarly be the probability distribution for the score $s_B$ of team $B$.  The probability of a victory by team $A$ is given by\footnote{Technically, this equation yields only the probability of an outright victory, $s_A > s_B$, without considering ties and overtime situations.}
\be 
\textrm{Prob}(s_A > s_B) = \int_0^\infty \int_{s_B + 1}^\infty P_B(s_B) P_A(s_A) ds_A ds_B.
\label{eq:winint}
\ee 
One can think of all strategizing by team $A$ as an attempt to choose its probability distribution $P_A(s_A)$ in such a way that Eq.\ (\ref{eq:winint}) is maximized.  

One can notice, of course, that the optimal strategy for $A$ requires one to know the strategy for $B$, as reflected by the probability distribution $P_B(s_B)$.  But team $B$ is similarly engaged in strategic decision making that requires knowledge about the strategy of $A$.  Thus, in general all strategic decisions are game-theoretical in nature and can in principle require complicated analytical descriptions.  In practice, however, the game-theoretical equilibria are often fairly obvious -- each team can use its opponent's previous record as an indicator of their intended strategy and ability, and adopt their own strategy accordingly.  The degree to which this approach can be expected to succeed, or can be exploited by a clever opponent, has not been explored in a rigorous way, but may prove a fascinating topic for future studies.

In general, Eq.\ (\ref{eq:winint}) is the quantity which is to be optimized by team $A$.  However, as mentioned in the Introduction, if one assumes that all possessions are independent of each other, then this description can be simplified by writing only a probability density function for the differential score $\Delta$.  The description becomes particularly simple in the limit of many remaining possessions, for which the distribution becomes Gaussian:
\be 
P(\Delta) = \frac{1}{\sqrt{2 \pi \sigma_\Delta^2}} e^{-(\Delta - \mu_{\Delta})^2 / 2 \sigma_\Delta^2 }.
\ee 
Here, the symbols $\mu_{\Delta}$ and $\sigma_{\Delta}^2$ denote the expected value and variance of the differential score.  These are related to the values for each team by 
\begin{eqnarray} 
\mu_{\Delta} & = & \mu_A - \mu_B, \\
\sigma_{\Delta}^2 & = & \sigma_{A}^2 + \sigma_{B}^2.
\end{eqnarray}

In this simplified description, the probability of a victory by team $A$ is $\mathcal{P} = \int_0^\infty P(\Delta) d\Delta$. 
This function increases monotonically with the parameter
\be 
Z = \frac{\mu_\Delta}{\sigma_\Delta} = \frac{\mu_A - \mu_B}{\sqrt{\sigma^2_A + \sigma^2_B}}.
\label{eq:Z}
\ee 
Therefore, in the limit where there are many possessions remaining in the game, one can express the problem of optimal strategy in the following simple way \citep{skinner_scoring_2011}:
\begin{center}
{\bf \textit{A team's optimal strategy is that which maximizes Z.}}
\end{center}

In general, $Z$ can be increased either by increasing the expected number of points $\mu_A$ of the team's strategy, or by altering the variance $\sigma_A^2$ of the team's strategy.  In particular, when $\mu_A - \mu_B$ is negative (team $A$ is the underdog), then $Z$ is increased by increasing the variance $\sigma^2_A$.  This is the mathematical formulation of the intuitive idea that underdogs must be willing to accept larger than average risk.  Similarly, when $\mu_A - \mu_B$ is positive, $Z$ is increased by \emph{reducing} the variance -- favored teams should play conservatively.  Thus, the example from the beginning of this section can be understood in terms of Eq.\ (\ref{eq:Z}).  

Most of the discussion surrounding the performance of actual basketball teams concerns efforts to increase a team's expected scoring ($\mu_A$) or to decrease their opponents' expected scoring ($\mu_B$).  Nonetheless, there are a variety of strategies designed primarily to alter the variance of the outcome [the denominator of Eq.\ (\ref{eq:Z}) rather than the numerator].  The most common technique is the choice of shot selection -- in particular, the choice to either increase (when trailing) or decrease (when leading) the proportion of $3$-point shots taken.  Another way to alter the variance of the outcome is by judicious manipulation of the game clock -- either by calling timeouts or by intentionally using the full length of a possession -- in order to manipulate the number of possessions remaining in the game.  Intentional fouling (``Hack a Shaq") can also be a useful way to alter the variance of the outcome, either by intentionally shortening possessions or by replacing potentially high-variance $3$-point opportunities with low-variance free throws.

Some elements of the reasoning outlined in this section have been appreciated in sports for a long time.  For example, the statistician Bill James developed as early as forty years ago a heuristic rule for when a lead in college basketball is ``safe" \citep{james_lead_2008}.  His rule, taken in the limit when there is time remaining for $N \gg 1$ possessions, is that a lead is safe when the lead is larger than the square root of the number of seconds remaining in the game.  The origin of this rule can be understood by examining Eq.\ (\ref{eq:Z}).    Since the probability of a victory by the leading team depends only on $Z$, then the probability that a lead is ``safe" is equivalent to the condition $Z > Z_c$, where $Z_c$ is a constant number.  For example, if you define ``safe" to mean a $98\%$ chance of victory, then $Z_c = 2.1$.  If one takes $\mu_\Delta$ to be equal to the current value of the lead $\Delta$, then the condition for a safe lead becomes $\Delta > \textrm{const.} \times \sqrt{\sigma^2_\Delta}$.  As discussed above, the variance $\sigma^2_\Delta$ of the outcome is directly proportional to the number of possessions remaining, and therefore it is proportional to the amount of time $\tau$ left in the game.  So the condition $Z > Z_c$ is equivalent to Bill James's heuristic condition for a safe lead: $\Delta > \textrm{const.} \times \sqrt{\tau}$.  This rule has since been studied carefully and refined in \citet{clauset_safe_2015}.

Other authors have made empirical studies of optimal strategy in specific scenarios that fall within the risk/reward paradigm.  For example, \citet{annis_optimal_2006} studied situations where the defensive team is protecting a three point lead in the final seconds of a basketball game, and found that the optimal strategy is generally to foul intentionally (thereby giving up two free throws) rather than allow for the possibility of a $3$-point shot by their opponent.  In the language of this section, one can say that this optimal strategy involves sacrificing points from the mean differential score in order to reduce variance in the score.  In \citet{wiens_crash_2013}, the authors studied the question of how much effort an offensive team should put into offensive rebounding.  Sending multiple players to ``crash the boards" for an offensive rebound has a potentially large reward, but comes with the risk of a poorly-defended fast break by the opponent.  The study quantified the costs and benefits of different offensive rebounding stragies, and reached the tentative conclusion that ``risky" offensive rebounding is usually more valuable than the more conservative strategy of hedging against fast breaks.

In \citet{goldman_misperception_2014}, the authors studied the risk/reward choices of NBA players, focusing on the frequency of $3$-point shots as an indicator of risk preference.  They found that players correctly respond to trailing situations with increased preference for risk, but also (incorrectly) exhibit increased preference for risk when their team has a large lead.  In this sense the general concepts of risk/reward tradeoff are not strongly ingrained in the habits of professional basketball players.

\section{Conclusion}
\label{sec:conclusion}
To summarize, this chapter has presented an analytical discussion of the different facets associated with optimal strategy in basketball.  Our primary message is that an optimal strategy must satisfy three conditions:
\begin{enumerate}
\item Allocative efficiency: All offensive options must have the same marginal efficiency.
\item Dynamic efficiency: The value of a shot taken at any instant in time must exceed the continuation value of the possession.
\item Risk/Reward tradeoff: An optimum strategy maximizes  the probability of victory, and not just the expectation value of the differential score.
\end{enumerate}
Methods for evaluating these criteria, and the degree to which they are met by actual basketball teams, have been discussed in the text in the references therein.

It is worth emphasizing, of course, that our results have been largely predicated on the assumption that individual possessions in a basketball game are IID.  The validity of this assumption has been a topic of some controversy in the literature for more than thirty years.  Most studies have focused on the existence of the ``hot hand" in basketball \citep{gilovich_hot_1985, wardrop_simpsons_1995, yaari_hot_2011, arkes_revisiting_2010, csapo_hand_2014, bocskocsky_heat_2014,  miller_surprised_2015}, a purported positive correlation between successive shots taken by a given player.  In fact there is some evidence for such a correlation \citep{arkes_revisiting_2010, yaari_hot_2011}, and other studies have found additional violations of the IID assumption, such as a decline in shooting effectiveness in high-pressure situations \citep{goldman_how_2013}, and an increase immediately following timeouts \citep{saavedra_is_2012}. There is also some evidence for a ``momentum effect", in which the probability of a win depends on the outcome of recent previous games \citep{arkes_finally_2011}.  Nonetheless, these correlations are generally very small in magnitude, so that any attempts to account for them in strategic decision making would likely be swamped by statistical uncertainty in other variables.

At a practical level, the biggest hindrance to quantitative basketball strategy is usually the difficulty of accurately estimating the efficiency of different offensive options.  The usage curves $f(p)$ are particularly difficult to estimate from easily-measurable statistics, and are necessary for a quantitative determination of optimal allocation.  What's more, the usage curves are really only robustly defined relative to a particular defense, and can vary strongly depending on the quality of the team's opponent.  A major advance in their determination may therefore provide the most important step toward enabling quantitative optimization of basketball strategy.  

It is also worth noting that, while the majority of this chapter has focused on offensive choices, another major challenge is to create a succinct analytical description of basketball \emph{defense}.  The defense also faces choices about how to allocate its effort and resources, and in this sense the considerations outlined in this chapter can be recast in terms of the defense as well.  Unfortunately, to our knowledge the necessary concept of ``defensive usage curves" has not been explored with sufficient depth to enable a real analysis.  

Despite these challenges, the level of sophistication associated with describing and optimizing the performance of basketball teams has increased enormously in the past decade.  This growth has been driven largely by a revolution in the way that descriptive data is recorded and analyzed \citep{lowe_seven_2013}, which has enabled hundreds of quantitative studies of basketball teams and players that would have been impossible only a decade ago.\footnote{To say nothing of the many more studies that are done privately by analysts working for professional basketball teams.}  We fully expect that within the coming decades many more important ideas will be developed that can be added to the ones presented here.

\bibliographystyle{apalike}
\bibliography{bibliography}


\end{document}